\documentclass[11pt]{article}

\usepackage{tabularx}
\usepackage{booktabs}
\usepackage{colortbl}
\usepackage{textcomp} % text version of symbols in TS1, e.g. \textrightarrow{}
\usepackage{amstext} % proper support of frequently used \text{} command
\newif\ifshowcitations\showcitationsfalse%
\newif\ifshowlinks\showlinksfalse%
\usepackage[numbers,sort&compress]{natbib}
\usepackage{graphicx}
\usepackage{amsmath}
\usepackage{dcolumn}
\usepackage{bm}
\usepackage{slashed}
\usepackage[bookmarks,bookmarksnumbered,colorlinks=true,anchorcolor=blue,
linkcolor=blue,urlcolor=blue,citecolor=blue,breaklinks=true]{hyperref}
\usepackage[utf8]{inputenc}
\usepackage{authblk}

\usepackage{color}
\usepackage{ulem}

\newcommand{\be}{\begin{equation}}
\newcommand{\ee}{\end{equation}}
\newcommand{\ba}{\begin{eqnarray}}
\newcommand{\ea}{\end{eqnarray}}
\def\bea{\begin{eqnarray}}
\def\eea{\end{eqnarray}}
\newcommand{\gsim}{\mathrel{\hbox{\rlap{\lower.55ex \hbox {$\sim$}}
                   \kern-.3em \raise.4ex \hbox{$>$}}}}
\newcommand{\lsim}{\mathrel{\hbox{\rlap{\lower.55ex \hbox {$\sim$}}
                   \kern-.3em \raise.4ex \hbox{$<$}}}}

\def\roughly#1{\mathrel{\raise.3ex\hbox{$#1$\kern-.75em%
\lower1ex\hbox{$\sim$}}}}
\def\lsim{\roughly<}
\def\gsim{\roughly>}

\def\({\left(}
\def\){\right)}
\def\[{\left[}
\def\]{\right]}
\def\<{\langle}
\def\>{\rangle}

\def\a{{\alpha}}

\usepackage{xcolor}

\usepackage{xcolor}

\title{\bf {The influence of Wilson lines on heavy quark anti-quark potential and mass}}

\author[1]{Bing Chen}
\author[1,2]{Xun Chen\thanks{chenxunhep@qq.com}}
\author[1]{Mitsutoshi Fujita\thanks{fujitamitsutoshi@usc.edu.cn}}
\author[3]{Jun Zhang\thanks{jzhang163@crimson.ua.edu}}
\affil[1]{School of Nuclear Science and Technology, University of South China, Hengyang 421001, China}
\affil[2]{INFN– Istituto Nazionale di Fisica Nucleare– Sezione di Bari Via Orabona 4, 70125, Bari, Italy}
\affil[3]{Department of Physics and Astronomy, University of Alabama, 514 University Boulevard, Tuscaloosa, AL, 35487, USA}
\renewcommand\Authands{ and }

\date{}

\begin{document}

\maketitle
%\vspace{0.1in}

\begin{abstract}
The holographic heavy quark potential is investigated via holographic Wilson loops in the AdS soliton with gauge potential. We analyze two types of holographic Wilson loops. 
{In the first type, holographic heavy quark potential shows the area law behavior. In the second type,  the potential becomes zero at a critical length and physics analogous to the dissociation occurs. The mass of heavy quarkonia and the binding energy are examined.} Lastly, the mass of $0^{++}$ glueball-like operators dual to massless dilaton is calculated. The mass of $0^{++}$ glueball-like operator decreases with increase of the gauge potential as expected in arXiv:2309.03491 [hep-th]. The results are comparable with lattice QCD.

\end{abstract}
{}

\newpage

\allowdisplaybreaks

\flushbottom

\section{Introduction}
The Wilson loop of a rectangular contour can be used to represent the quark-antiquark potential. The contour is characterized by a rectangle with length $\tau$ in the temporal direction and width $L$ in the spatial direction, where $\tau \gg L$. The parameter $L$ denotes the separation between the quark and antiquark pair. The expectation value of the Wilson loop is given by:
\begin{align}
\langle W(C) \rangle \sim e^{-\tau V(L)},
\end{align}
where $V(L)$ denotes the quark-antiquark potential.

The Wilson loop exhibits an area law behavior in the confinement regime, where $\tau V(L) \propto \tau L$. This contrasts with the Coulomb-like behavior $V(L) \propto 1/L$ observed in perturbative QCD, highlighting the inadequacy of perturbative QCD at low-energy scales.

The holographic quark-antiquark potential has been analyzed using the minimal surface of the string~\cite{Maldacena:1998im, Rey:1998ik}. The area law behavior of the holographic Wilson loop, signifying confinement, was confirmed in~\cite{Witten:1998zw, Brandhuber:1998er}. The holographic quark-antiquark potential at finite temperature has been studied in~\cite{Rey:1998bq, Brandhuber:1998bs,Bak:2007fk}, where it exhibits Coulomb-like behavior with a finite range and shows the dissociation of heavy quarkonium beyond a critical length. This dissociation is a key signal of deconfinement. The potential energy between quarks and antiquarks is crucial for understanding hadron formation in bound states. By examining the interquark potential, insights into the structure of hadrons and the underlying dynamics of QCD can be gained. The quark-antiquark potential under extreme conditions (e.g., magnetic fields, rotation, etc.) has been studied through holography in Refs.~\cite{Andreev:2006nw, Mia:2010zu, Dudal:2017max, Chen:2017lsf, Chen:2020ath, Guo:2024qiq, Zhu:2024dwx, Luo:2024iwf, Shi:2021qri, Li:2011hp, Wu:2024sdl, Zhang:2015faa, He:2010ye, Yang:2015aia, Fadafan:2012qy, Deng:2021kyd, Chen:2024edy, Arefeva:2023jjh, Bohra:2019ebj, Colangelo:2010pe}. This research aims to explore the correlation characteristics of quarkonium pairs through the analysis of the quark potential. In particular, the mass of heavy quarkonium and its binding energy in the nonrelativistic limit are consistent with experimental data. Investigating the dependence of quarkonium mass on (imaginary) chemical potential or magnetic fields presents an intriguing direction for further study.

{To analyze heavy quarkonium, we have expanded a bottom-up model that has previously served as a holographic model for color superconductivity~\cite{Basu:2011yg, Ghoroku:2019trx}. Owing to the QCD scale or confinement, the gravity dual incorporates an extra scale. AdS solitons possess an IR scale originating from the radius of the compactified direction, which corresponds to a confined phase with a mass gap. This leads to the emergence of a discrete spectrum of glueball states in the dual field theory~\cite{Csaki:1998qr}. The AdS soliton exhibits negative energy~\cite{Horowitz:1998ha} and remains stable against perturbations in any dimensions, in line with the positive energy conjecture~\cite{Constable:1999gb}. This negative energy stems from the anti-periodic boundary condition imposed on fermions, which shatters supersymmetry. In the context of Yang-Mills theory on $S^1\times R^{1,2}$  with twisted boundary conditions, the Casimir energy is also negative, barring the extremal limit where the energy dwindles to zero, contingent upon the twist parameter~\cite{Nishioka:2006gr}. The twisting parameter concurrently alters the boundary condition and Casimir energy of a spin field in two dimensions~\cite{Polchinski:1998rq}. Specifically, it takes on a negative value for the anti-periodic boundary condition and a positive value for the periodic boundary condition. Hence, the twist parameter is a fascinating topic for investigation. } 

%The holographic quark-antiquark potential was analyzed by using the minimal surface of the string~\cite{Maldacena:1998im,Rey:1998ik}.  The area law behavior of the holographic Wilson loop has been confirmed in~\cite{Witten:1998zw,Brandhuber:1998er}, which describes confinement. The holographic quark-antiquark potential was also analyzed at finite temperature in~\cite{Rey:1998bq,Brandhuber:1998bs}. \textcolor{blue}{The dissociation of heavy quark-antiquark pairs is a key signal of deconfinement, which is of great interest because the potential energy between quarks and antiquarks is essential for understanding the formation of hadrons in bound states. By examining this interquark potential, researchers can gain insights into the structure of hadrons and the underlying dynamics of QCD. Thus, the goal of this research is to explore the correlation characteristics of quarkonium pairs through the analysis of the quark potential.} Especially, the mass of heavy quarkonium and binding energy in the nonrelativistic limit are comparable with experimental datas. The dependence of the mass on (imaginary) chemical potential or the magnetic field is an interesting direction.

Another motivation for this study is to determine the mass of operators in the QFT dual to the AdS soliton with a gauge potential (a twisting parameter), as the operator mass is crucial for analyzing degrees of freedom. This background under consideration arises from a double Wick rotation of the Reissner-Nordström AdS black hole with an imaginary chemical potential (see \cite{Ghoroku:2020fkv} for the analytic continuation of the Reissner-Nordström AdS black hole). Since the imaginary chemical potential corresponds to the twist parameter after a gauge transformation, the gauge potential can be interpreted as introducing a twisted boundary condition along the circle in the cigar geometry. The gauge potential affects the Casimir energy of the dual QFT, rendering it positive, which implies an increase in the degrees of freedom. Additionally, Wilson lines shift the masses of charged particles, a phenomenon also observed in string theory~\cite{Polchinski:1998rq} (see Appendix \ref{appA}). These mass shifts are reflected in the entropic $C$-function and the renormalized entanglement entropy, both of which quantify degrees of freedom~\cite{Fujita:2020qvp, Fujita:2023bdk}. Specifically, as the operator mass increases, these quantities decrease significantly in the large $l$ limit, indicating the decoupling of massive degrees of freedom. Thus, exploring the mass of operators in the dual QFT remains an intriguing and valuable direction for further research.

%Another motivation is to obtain the mass of operators in QFT dual to the AdS soliton with gauge potential because mass of operators is relevant for analyzing degrees of freedom. This background is the double Wick rotation of the Reissner Nordstr\"{o}m AdS black hole with imaginary chemical potential (see \cite{} for the analytic continuation of the Reissner Nordstr\"{o}m AdS black hole). Because imaginary chemical potential corresponds to the twist parameter, the interpretation of gauge potential is to include a twisted boundary condition along a circle of the cigar direction.  Gauge potential can change Casimir energy of dual QFT and makes it positive. It implies that gauge potential increases degrees of freedom. Moreover, Wilson lines shift the mass of charged particles. This shift is also seen in string theory~\cite{Polchinski:1998rq} (see appendix \ref{appA}). The alternations of mass were captured by the entropic C function and the renormalized entanglement entropy, which represent degrees of freedom~\cite{Fujita:2020qvp,Fujita:2023bdk}. That is, as the operator mass increases, these quantities decrease well in large $l$. It implies the decoupling of massive degrees of freedom. Thus, it will be interesting to explore the mass of operators in dual QFT.

This paper examines the quark-antiquark potential and the mass spectrum of a glueball-like operator in the AdS soliton background with a gauge potential. We extract the quark-antiquark potential from the holographic Wilson loop and the associated area law. Given the connection between the string tension and the mass of excited modes, we investigate the dependence of the string tension on the gauge potential. {Additionally, another type of holographic quark-antiquark potential demonstrates the dissociation of heavy quarkonium. We analyze heavy quarkonium's binding energy and mass from the Schr{\"o}dinger equation.} The holographic results are comparable to  Bottomonium. Bottomonium is a bound state composed of heavy bottom quarks carrying charge -1/3, specifically denoted as, $\Upsilon =b\bar{b} $~\cite{Eichten:1978tg,Eichten:1979ms,Perkins:1982xb}. It is a relevant state for analyzing the non-relativistic Schr{\"o}dinger equation.
Moreover, we examine the spectrum of the glueball-like operator in the presence of the gauge potential. This operator is a four-dimensional scalar, originally arising as the supersymmetric completion of the glueball operator in $\mathcal{N}=4$ SYM theory in four dimensions.

This paper is organized as follows. In Section 2, we review the AdS soliton with a gauge potential, analyze the total energy of the spacetime {, and discuss degrees of freedom on the QFT side.} In Section 3, we investigate the holographic heavy quark potential derived from the minimal surface of strings. {We also employ the Schr{\"o}dinger equation to calculate the masses of the excitations and binding energy}. In Section 4, we focus on the massless dilaton dual to the glueball-like operator and compare the spectrum of this operator with lattice QCD data.

%In this paper, we analyze the holographic Wilson loop and mass spectrum of glueball-like operator in the AdS soliton with gauge potential. We derive the quark-antiquark potential from the holographic Wilson loop and the area law. Because the string tension is related to the mass of excited modes, we are interested in gauge potential dependence of the string tension. We also analyze the spectrum of the glueball-like operator in the presence of gauge potential. This is a 4 dimensional scalar operator, which is originally supersymmetric completion of the glueball operator in $\mathcal{N}=4$ SYM theory in 4 dimensions.

%This paper is organized as follows. In section 2, we review the AdS soliton with gauge potential. We analyze total energy of spacetime. In section 3, we analyze holographic heavy quark potential from the minimal surface of strings. We derive the mass of the excitations from heavy quark potential. In section 4, we analyze massless dilaton dual to glueball-like operator. The spectrum of the glueball-like operator is compared with datas of lattice QCD.

\section{Spacetime}
{In this section, we review the AdS soliton with a gauge potential. When the gauge potential is switched off, the AdS soliton is regarded as being dual to a confining gauge theory. This theory exhibits confinement, a mass gap, and a discrete mass spectrum of excitations, which are common features in our setups. For D3 and M5 branes, the dual theory contains the glueballs of QCD$_3$ and QCD$_4$, respectively~\cite{Constable:1999gb}. QCD$_3$ can be seen as the high-temperature limit of $4d$ Super Yang-Mills theory, essentially representing $3d$ pure Yang-Mills theory at long distances~\cite{Witten:1998zw}. Similarly, a $4d$ gauge theory appearing in this paper corresponds to the high-temperature limit of a $5d$ gauge theory at an imaginary chemical potential and at long distances dual to the Reissner-Nordström AdS black hole.}

{The analytic continuation of the Reissner Nordstr\"{o}m AdS black hole was studied in \cite{Ghoroku:2020fkv} to model a QCD-like system with an imaginary chemical potential.  Substituting $\mu =ia_{\phi}$ and $t= -i\tau$ into the metric of the 6-dimensional Reissner Nordstr\"{o}m AdS black hole ($d=5$)~\cite{Hartnoll:2009sz}, we arrive at
\ba
&ds^2=\dfrac{R^2}{z^2}\Big(f(z)d\tau^2 +\dfrac{dz^2}{f(z)}+dx_ldx_l\Big), \nonumber \\
&A_{\tau}=a_{\phi}^{(0)}\Big(1-\Big(\dfrac{z}{z_+}\Big)^{d-2}\Big),
\ea
where $f(z)=1-(1-z_+^2a_{\phi}^{(0)2}/\delta^2)(z/z_+)^d-z_+^2a_{\phi}^{(0)2}/\delta^2 (z/z_+)^{2(d-1)}$. {$\delta$ is a factor $\delta^2 =(d-1)g^2R^2/((d-2)\kappa^2)$, where $g$ and $\kappa$ are coupling constants.} {Adding antisymmetric tensor $B$, the holographic model shows the Roberge-Weiss periodicity at the imaginary chemical potential~\cite{Roberge:1986mm,Aarts:2010ky}. A lower bound of the color number $N_c$ for the Roberge-Weiss periodicity was obtained in \cite{Ghoroku:2020fkv}: $N_c\ge \sqrt{15}/2\delta \sim 1.2$. 
To see dependence on a twisting parameter clearly, we currently omit the antisymmetric tensor $B$.}

%The analytic continuation of the Reissner Nordstr\"{o}m AdS black hole was considered in \cite{} to analyze a system like QCD with imaginary chemical potential (see~\cite{Hartnoll:2009sz} for the metric of the Reissner Nordstr\"{o}m AdS black hole). This holographic model shows the Roberge-Weiss periodicity at the imaginary chemical potential. 

{By applying the analytic continuation of spacetime coordinates, the metric of the double Wick rotated Reissner Nordstr\"{o}m AdS black hole ($\tau \leftrightarrow \phi$) with imaginary chemical potential becomes
\ba\label{MET01}
&ds^2=g_{\mu\nu}dx^{\mu}dx^{\nu}=\dfrac{R^2}{z^2}\Big(\dfrac{dz^2}{f(z)}+f(z)d\phi^2+d\tau^2+dx_idx_i\Big), \nonumber \\
& A_{\phi}=a_{\phi}^{(0)}\Big(1-\Big(\dfrac{z}{z_+}\Big)^{d-2}\Big),
\ea
where $\tau$ is Euclidean time and $f(z)=1-(1-z_+^2a_{\phi}^2)(z/z_+)^d-z_+^2a_{\phi}^2 (z/z_+)^{2(d-1)}$ ($a_{\phi}=a_{\phi}^{(0)}/\delta$).} See papers~\cite{Fujita:2020qvp,Fujita:2023bdk}. The radial coordinate is bounded by $z\le z_+$, and the $\phi$ direction exhibits periodicity $\phi\to \phi +1/M_0$ due to the regularity at the tip. The KK mass becomes
\ba
M_0=\dfrac{1}{4\pi z_+}\Big(d+(d-2)a_{\phi}^2z_+^2\Big).
\ea
This formula can be rewritten as
\ba\label{ZP3}
z_+=\dfrac{d}{2\pi M_0\pm \sqrt{4\pi^2 M_0^2-d (d-2)a_{\phi}^2}}.
\ea
There are two branches. The plus sign must be chosen because the metric approaches the AdS soliton in small $a_{\phi}$ limit. According to~\cite{Fujita:2020qvp}, the solution with the plus sign is more stable. 
Equation \eqref{ZP3} indicates that the solitonic solution only exists for $a_{\phi}\le 2\pi M_0/\sqrt{d (d-2)}$.
The circle along the $\phi$ direction is small for large KK mass. The dual theory is described by $d-1$ dimensional field theory at low energy.

\subsection{The boundary energy}
The energy of the AdS soliton with gauge potential can be derived from the total energy of spacetime~\cite{Brown:1992br}. This background possesses time translation symmetry. The total energy is defined as
\ba\label{TOT39}
M=\dfrac{1}{\kappa^2}\int d^{d-1}x N \sqrt{\sigma} (K-K_0).
\ea
The integral is evaluated on a spatial slice at constant $t$ and fixed $z$. The extrinsic curvature $K_0$ is evaluated on the reference spacetime ($f(z)=1$). One then has the area element $A=\sqrt{f(z)}(R/z)^{d-1}V_{d-2}/M_0$, where  $V_{d-2}$ is the volume of $d-2$ dimensional spacetime. The lapse function is defined as $N=R/z$. Consequently, \eqref{TOT39} is evaluated as 
\ba
M=-\dfrac{V_{d-2}}{M_0}\frac{R^{d-1}}{2\kappa^2 z_+^d }\Big(1-{z^2_+ a^2_\phi}\Big).
\ea

This energy is equivalent to the boundary energy computed with the help of counterterms~\cite{Henningson:1998gx, de Haro:2000xn, Balasubramanian:1999re}. The latter is expressed as a volume integral of the expectation value of the stress tensor, as detailed in~\cite{Fujita:2020qvp}:
\ba\label{Ttt1}
&\langle T^{(0)}_{tt}\rangle=-\dfrac{R^{d-1}}{2\kappa^2}\dfrac{1}{z^d_+}\Big(1-{z^2_+ a^2_\phi}\Big), \nonumber \\
&\int d\phi d^{d-2}x \langle T_{tt}^{(0)}\rangle =M.
\ea

The boundary energy can increase with an increase in the gauge potential as follows: \ba
 \begin{cases}\label{MBD7}
M<0 \quad \text{$a_{\phi} <\dfrac{2\pi M_0}{d-1}$} \\
M>0  \quad \text{$a_{\phi} >\dfrac{2\pi M_0}{d-1}$}.
 \end{cases}
 \ea
 When $a_{\phi}=2\pi M_0/(d-1)$, boundary energy vanishes. This behavior will be analogous to the extremal limit of the Casimir energy in the Yang-Mills theory on $S^1\times R^{1,2}$ with twisted boundary conditions~\cite{Nishioka:2006gr}. In this context, the Casimir energy is negative except at the extremal limit, where energy vanishes. For $a_{\phi}=0$ and $d=4$, \eqref{MBD7} realizes Casimir energy of $4d$ SYM theory on $R^3\times S^1$~\cite{Horowitz:1998ha}.

\section{{The holographic Wilson loop and heavy quark potential}}
\subsection{$\phi= constant $}
We investigate holographic quark anti-quark potential from the holographic Wilson loop in the AdS soliton, including a gauge potential. This potential is obtained from the minimal surface of the string~\cite{Maldacena:1998im, Rey:1998bq}.
We consider a worldsheet spanned by the coordinates $\tau$ and $x^1=x$ ($-L/2\le x \le L/2$). The $\tau$ direction is chosen to be infinitely long. The system is translationally invariant along this direction.  The static metric is \eqref{MET01}, where $R$ is set to be a constant 1. {The heavy quark potential is given by the Namb-Goto action, which captures the low-energy dynamics of the string as follows~\cite{Rey:1998bq}:}
 {\ba\label{ENE8}
 E=\frac{S }{\tau},\quad S=S_{NG}={\frac{1}{2\pi \alpha'}} \int d\xi^0 d\xi^1 \sqrt{-\det g_{\alpha\beta}}
 \ea}
 where $\alpha'$ is a constant 1 , $(\xi_0, \xi_1)$ denote worldsheet coordinates, and $g_{\alpha\beta}$ represents an induced metric.  {We adopt the static gauge where $\xi_0 = \tau, \xi_1 = x$, and treat the string coordinate $z$ as a function of $x$ to identify static string configurations representing a pair of heavy quark and anti-quark. }
 The coordinate $\tau$ also signifies the time separation. The boundary condition for the function $z(x)$ is
 \ba\label{BDY1}
 z(\pm \frac{L}{2})= 0, \quad z(0) = z_0, \quad \left(\partial_x z\right)^2 \Big|_{z=z_0} = 0.
 \ea
{The boundary conditions \eqref{BDY1} are pertinent to a U-shaped string connecting quark and anti-quark.  The first one indicates that the endpoint of the string is at the AdS boundary at $x=\pm L/2$, where $L$ is {the inter-quark distance}. }
 
Upon substituting the metric \eqref{MET01}, the action $ S $ is expressed as
\ba\label{ACT10}
S = \frac{1}{2\pi \alpha'}\int d\tau \int dx\sqrt{g_{\tau\tau}(g_{xx} +g_{zz} z'^2)}=\frac{1}{2\pi \alpha'} \int d\tau \int R^2\sqrt{\frac{1}{z^2}(\frac{1}{z^2}+\frac{z'^2}{z^2 f(z)})}dx .
\ea

Now we define the Lagrangian as:
\ba
\mathcal{L} = R^2\sqrt{\frac{1}{z^2}(\frac{1}{z^2}+\frac{z'^2}{z^2 f(z)})},
\ea
Since $ \mathcal{L} $ does not depend on the coordinate $ x $ explicitly, we have the conserved quantity (the Hamiltonian),
\ba
\mathcal{L} - \frac{\partial\mathcal{L}}{\partial z'} z'.
\ea
When $z = z_0 $, $z'|_{z=z_0}=0$ and the conserved quantity simplifies to
\ba
\frac{1}{z^2\sqrt{(1+\frac{z'^2}{f(z)})}}=\frac{1}{z_0^2}
\ea
From the aforementioned equation, one derives the following first-order differential equation:
\ba\label{ZPR14}
z'=\sqrt{\frac{z_0^4}{z^4}f(z)-f(z)}
\ea
Subsequently, we establish the relationship between $z_0$ and {the inter-quark distance}  $L$ as follows:
\ba\label{LEN17}
L= 2 \int_0^{z_0} \frac{\partial x}{\partial z} \, dz.= 2 \int_0^{z_0} \frac{1}{\sqrt{\frac{z_0^4}{z^4}f(z)-f(z)}} dz,
\ea
where ${\partial_z x}= \frac{\partial x}{\partial z} = \frac{1}{\partial_x z}=\frac{1}{z'}$.

Substituting \eqref{ZPR14} into the action $S$ \eqref{ACT10}, we arrive at
\ba
S = \frac{\tau R^2}{\pi \alpha'}\int_0^{z_0}\sqrt{\frac{z_0^4}{z^4 f(z)(z_0^4-z^4)}}dz.
\ea
{This is the on-shell action of a U-shaped string with a turning point $z=z_0$. Based on \eqref{ENE8}, the regularized energy of the U-shaped string of the inter-quark distance $L$ is expressed as
\ba\label{ENE19}
E=\frac{ R^2}{\pi \alpha'}\int_0^{z_0}\Big(\sqrt{\frac{z_0^4}{z^4 f(z)(z_0^4-z^4)}}-\frac{1}{z^2\sqrt{f(z)}}\Big)dz-\dfrac{R^2}{\pi \alpha'}\int^{z_+}_{z_0}\dfrac{dz}{z^2\sqrt{f(z)}},
\ea
where the subtraction has carried out up to the soliton's tip~\cite{Brandhuber:1998bs}.}

\begin{figure}[htbp]
     \begin{center}
   \includegraphics[height=4cm,clip]{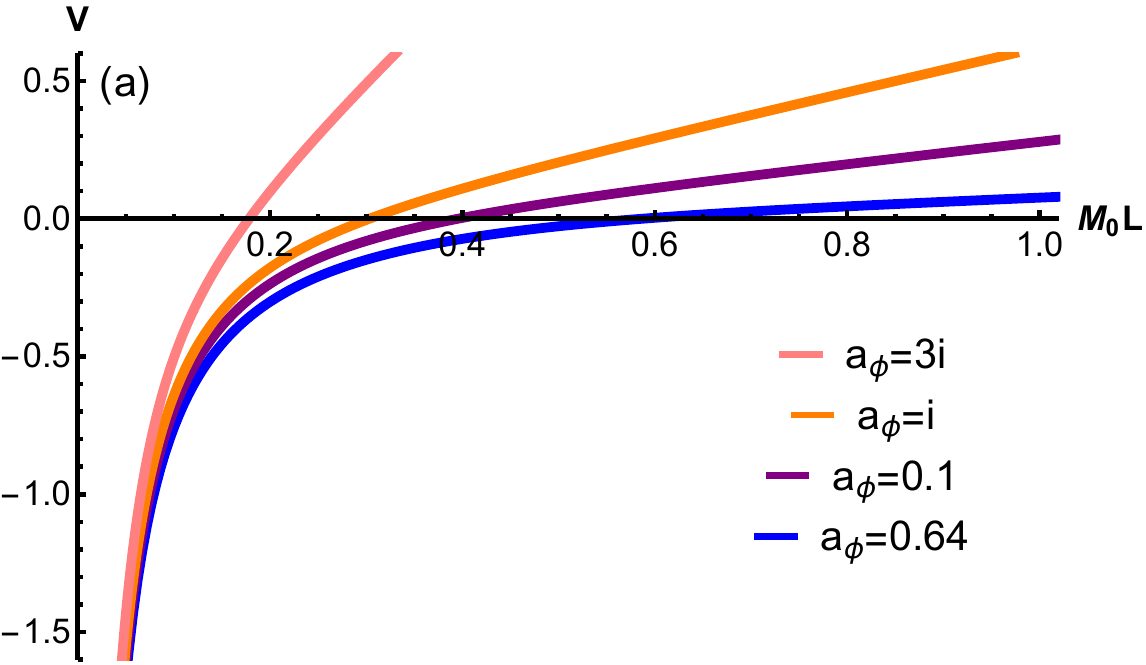}
   \includegraphics[height=4cm,clip]{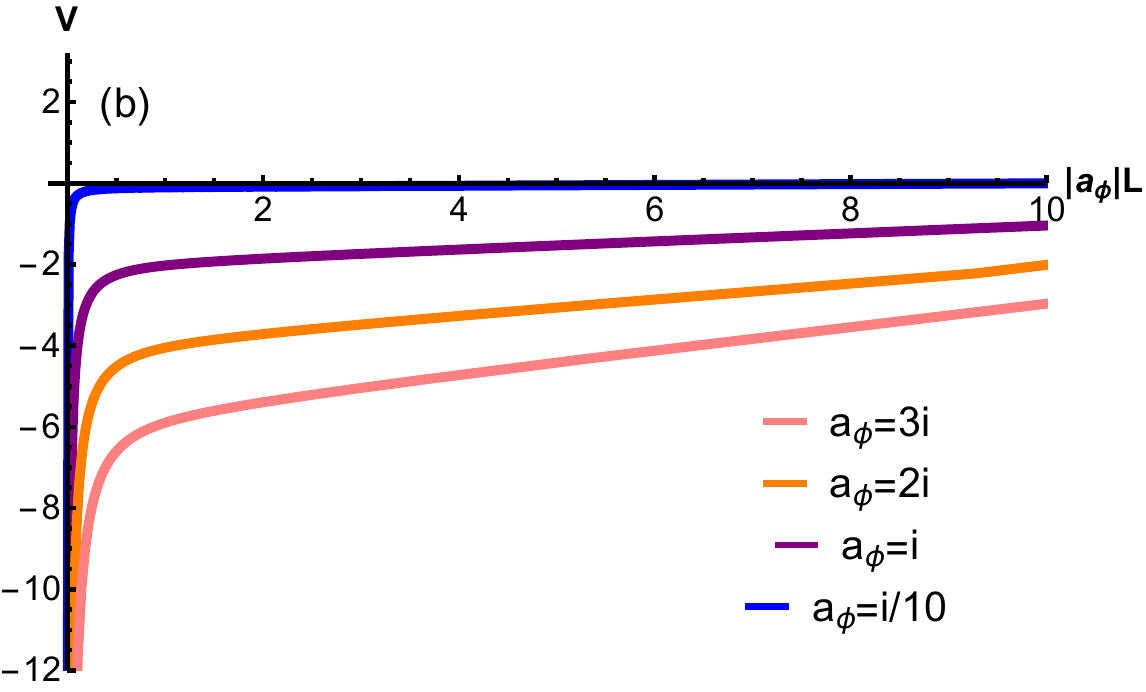}
           \caption{(a) The normalized quark anti-quark potential $V=\alpha'E/R^2$ is plotted as as a function of $M_0L$ ($M_0=0.4$). The gauge potential is changed. The area law behavior is observed in any case. The QCD string tension (represented by the slope) is suppressed when the gauge potential is large. (b) The quark anti-quark potential $V$ is shown as a function of $|a_{\phi}|L$ ($M_0=0$). Note that $a_{\phi}$ must be imaginary. The area law is observed except when $a_{\phi}=0$. The QCD string tension increases when $|a_{\phi}|$ increases.}
    \label{fig:Wil}
    \end{center}
\end{figure}
\begin{figure}[htbp]
     \begin{center}
   \includegraphics[height=5cm,clip]{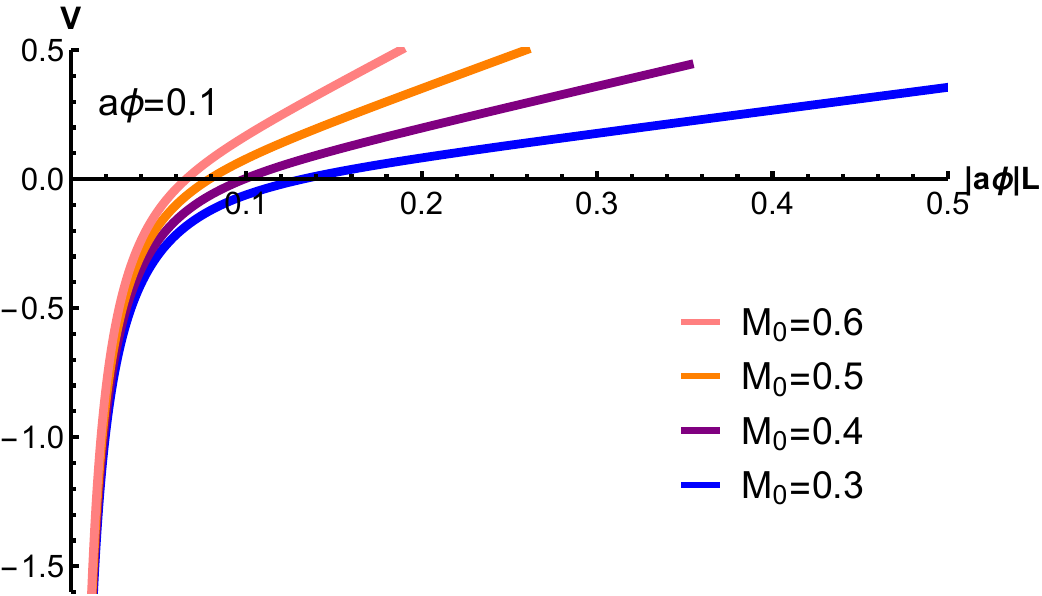}
           \caption{The normalized quark anti-quark potential is plotted as a function of $|a_{\phi}|L$ for $a_{\phi}=0.1$. The area law behavior is evident in any case. The QCD string tension (represented by the slope) increases with the Kaluza-Klein mass $M_0$ increase.
}
    \label{fig:Wil0}
    \end{center}
\end{figure}

In Fig. \ref{fig:Wil}(a), the normalized quark anti-quark potential $V=\alpha'E/R^2 $ is depicted as a function of $M_0L$ in $d=5$ dimensions. Because $V$ is directly proportional to {the inter-quark distance} $L$, we see the area law behavior at large $M_0L$. That indicates that the theory is in the confining phase. The slope is equal to the QCD string tension. The QCD tension diminishes with an increase of $a_{\phi}$. It implies confinement is less pronounced at large $a_{\phi}$. In Fig. \ref{fig:Wil}(b), the $d=5$ quark anti-quark potential $V$ is plotted as a function of $|a_{\phi}|L$ in an extremal limit ($M_0=0$). Due to scaling symmetry of the EOM under $(t,\vec{x},z)\to \lambda (t,\vec{x},z)$, there is the scaling transformation of parameters $(a_{\phi},V,1/L)\to  \lambda^{-1} (a_{\phi},V,1/L)$. Therefore, the potential has the functional form $V=|a_{\phi}|F(|a_{\phi}|L)$.
We again witness the area law at large $|a_{\phi}|L$, except for $a_{\phi}=0$, where the background becomes pure AdS. The absence of the area law $a_{\phi}=0$ is attributed to the massless nature of the excitation.

In Fig. \ref{fig:Wil0}, the normalized quark anti-quark potential is depicted as a $|a_{\phi}|L$ function. When $M_0$ increases, the QCD string tension, reflected in the potential slope, increases. It implies that the phenomenon of confinement is enhanced due to an area law.
\begin{figure}[htbp]
     \begin{center}
   \includegraphics[height=5cm,clip]{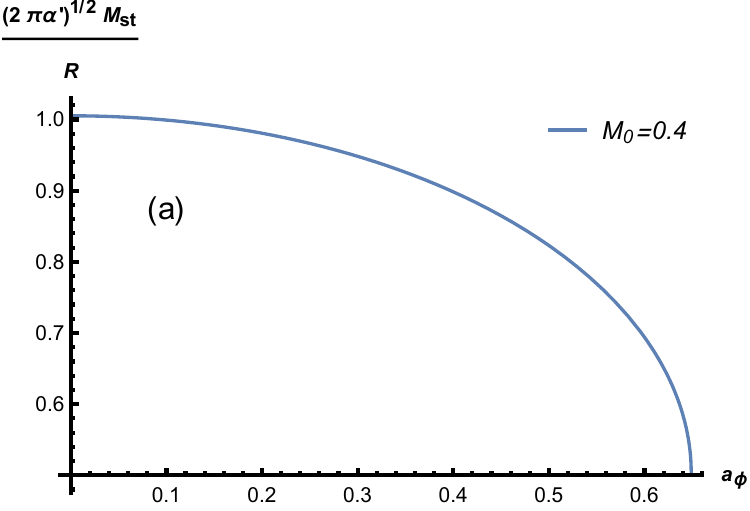}
   \includegraphics[height=5cm,clip]{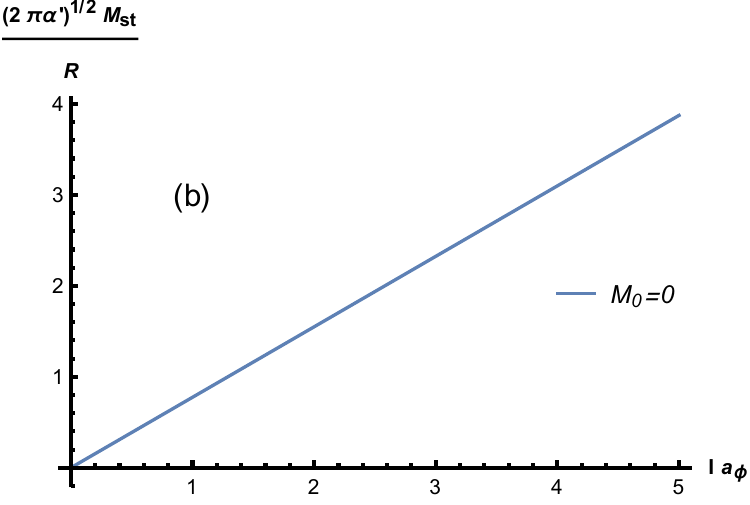}
           \caption{(a) The normalized mass of the excitation of QCD strings is plotted as a function of $a_{\phi}$ ($M_0=0.4$). The mass decreases as $a_{\phi}$ increases. More excited modes contribute to DOF at large $a_{\phi}$. (b) The normalized mass is plotted as a function of $| a_{\phi}|$ in the extremal limit ($M_0=0$). Note that $a_{\phi}$ is purely imaginary. For $a_{\phi}=0$, both $M_0$ and $M_{st}$ are zero, while $M_{st}$ increases as $|a_{\phi}|$ increases. It implies that the mass of the excitation decouples others in a slightly low energy limit when  $|a_{\phi}|$ increases.}
    \label{fig:Mst}
    \end{center}
\end{figure}

{Furthermore, our analysis also shows that a spatial Wilson loop in the high-temperature limit (large $M_0$) of a $5d$ gauge theory with an imaginary chemical potential obeys an area law. This gauge theory is dual to a 6-dimensional Reissner-Nordstr{\"o}m black hole. In addition, the resulting $4d$ gauge theory also obeys an area law (a confined phase) at long distances $(M_0L\gg 1)$. While an area law has been confirmed for cases such as the AdS soliton, proving the area law is important because the gravity dual is more unstable than an AdS black hole for a large gauge potential $a_{\phi}$. }

We perform an analytical study of the excitation of QCD strings and its $a_{\phi}$ dependence. Specifically, we take the limit $z_0\to z_+$. The dominant contribution of the integral arises from the region $z\sim z_+$. By employing equations \eqref{ENE19} and \eqref{LEN17}, energy is obtained as
\ba\label{ARE32}
E=T_{QCD} L,
\ea
where the QCD string tension becomes
\ba
T_{QCD}=\dfrac{ R^2}{2\pi \alpha ' z_+^2}=\dfrac{R^2}{2\pi\alpha'}\Big(\dfrac{2\pi M_0+\sqrt{4\pi^2 M_0^2-d(d-2)a_{\phi}^2}}{d}\Big)^2.
\ea
The QCD string tension is associated with the mass of the excitation of the QCD string. Specifically, this mass becomes $M_{st}=R/(z_+\sqrt{\alpha'})$ and scales linearly with $R$. For small values of $a_{\phi}$, $M_{st}\sim RM_0/\sqrt{\alpha'}$, which is extremely large in the supergravity limit $R/\sqrt{\alpha'}\gg 1$. Consequently, this mode is only accessible at high energies $E\gg M_0$. All Kaluza-Klein modes at such high energies are from the $\phi$ circle. Thus, \eqref{ARE32} elucidates the area law of higher dimensional theory, essentially $d$-dimensional QFT on a circle~\cite{Brandhuber:1998er}.

Fig. \ref{fig:Mst} (a) illustrates that $M_{st}$ diminishes as $a_{\phi}$ increases $(M_0=0.4)$. When the gauge potential is substantial $a_{\phi}=2\pi M_0/\sqrt{d(d-2)}$, the mass is reduced to half of $M_{st}(a_{\phi}=0)$. This indicates that the excitation mode contributes to DOF when $a_{\phi}$ is large because it does not cause other modes to decouple in a slightly low energy limit. Consequently, the area law is mitigated, and confinement is suppressed.
%We will see that confinement is also suppressed at large $a_{\phi}$ and the deconfinement phase becomes more likely.

Fig. \ref{fig:Mst} (b) demonstrates that $M_{st}$ increases as $| a_{\phi}|$ increases on an extremal case $(M_0=0)$. $a_{\phi}$ must be purely imaginary. This is intriguing because KK modes are massless in an extremal limit, while excited modes of QCD strings retain mass. Because KK modes are massless, they significantly contribute to DOF.
%However, excited modes cause other modes to decouple at large $a_{\phi}$ and slightly low energy limit due to their %increasing mass.

\subsection{$x= constant $}
This section examines holographic quark anti-quark potential along the $\phi$ direction.
 We take into account the worldsheet along with $\tau$ and $\phi$ ($-L/2\le \phi \le L/2$). $\tau$ direction is chosen to be infinitely long. The system maintains invariance under the translation along this direction. The static metric becomes \eqref{MET01}.
{ The holographic heavy quark potential is given by the following Namb-Goto action as the low-energy action of the string~\cite{Rey:1998bq}:}
 \ba
 E=\frac{S }{\tau},S=S_NG=\frac{1}{2\pi \alpha'} \int d\xi^0 d\xi^2 \sqrt{-\det g_{\alpha\beta}}
 \ea
where $\alpha'$ is a constant set to 1 , $(\xi_0, \xi_2)$ denote the worldsheet coordinates, and $g_{\alpha\beta}$ is an induced metric. {We adopt the static gauge where $\xi_0 = \tau, \xi_2 = \phi$, and treat the string coordinate $z$ as a function of $\phi$ for representing static string configurations of a pair of heavy quark and anti-quark.} 
 The boundary condition for $z(\phi)$ is
 \ba\label{ZPH1}
 z(\pm \frac{L}{2})= 0, \quad z(0) = z_0, \quad \left(\partial_\phi z\right)^2 \Big|_{z=z_0} = 0.
 \ea
{The boundary conditions \eqref{ZPH1} are useful for a U-shaped string connecting quark and anti-quark. The first indicates that the string's end point is at the AdS boundary ($x=\pm L/2$).}

Upon substituting the metric from \eqref{MET01}, the action $ S $ is transformed into:
\ba
S = \frac{1}{2\pi \alpha'}\int d\tau \int d\phi \sqrt{g_{\tau\tau}(g_{\phi\phi} +g_{zz} \dot{z}^2)}=\frac{1}{2\pi \alpha'} \int d\tau \int R^2\sqrt{\frac{1}{z^2}(\frac{f(z)}{z^2}+\frac{\dot{z}^2}{z^2 f(z)})}d\phi .
\ea
The Lagrangian is defined as
\ba
\mathcal{L} = R^2\sqrt{\frac{1}{z^2}(\frac{f(z)}{z^2}+\frac{\dot{z}^2}{z^2 f(z)})},
\ea
Because the Lagrangian $\mathcal{L}$ does not depend on $\phi$, one can have the conserved quantity. That is, the Hamiltonian is a constant. When $z = z_0$, the following formula is obtained
\ba\label{HAM12}
\frac{f(z)}{z^2\sqrt{(f(z)+\frac{z'^2}{f(z)})}}=\frac{f(z)}{z_0^2\sqrt{(f(z_0)}}
\ea
where the condition $z'|_{z=z_0}=0$ was utilized. Solving \eqref{HAM12} with respect to $z'$, one obtains the following 1st order differential equation:
\ba
\dot{z}=\sqrt{\frac{f(z)^3z_0^4-f(z)^2f(z_0)z^4}{z^4f(z_0)}}
\ea
{Additionally, the inter-quark distance $ L $ is calculated in the following manner:}
\ba\label{LEN31}
L = 2 \int_0^{z_0} \frac{\partial \phi}{\partial z} \, dz= 2 \int_0^{z_0} \frac{1}{\sqrt{\frac{f(z)^3z_0^4-f(z)^2f(z_0)z^4}{z^4f(z_0)}}} dz.
\ea
where ${\partial_z \phi}=( \frac{\partial z}{\partial \phi})^{-1} = \frac{1}{\partial_\phi z}=\frac{1}{\dot{z}}$.
Consequently, we have the potential after eliminating the infinite terms:
\ba\label{ACT29}
S = \frac{\tau R^2}{\pi \alpha'}\int_0^{z_0}\frac{z_0^2}{z^2}\sqrt{\frac{1}{z_0^4 f(z)-z^4f(z_0)}}dz
\ea
{After eliminating the infinite terms, the regularized energy of the U-shaped string of the inter-quark distance $L$ becomes
\ba\label{ENEN}
E=\frac{R^2}{\pi \alpha'}\int_0^{z_0}\Big(\frac{z_0^2}{z^2}\sqrt{\frac{1}{z_0^4 f(z)-z^4f(z_0)}}-\frac{1}{z^2\sqrt{f(z)}}\Big)dz-\dfrac{R^2}{\pi\alpha'}\int^{z_+}_{z_0}\dfrac{dz}{z^2\sqrt{f(z)}}.
\ea
Note that the string's worldsheet is oriented along the $(\tau,\phi)$ axes analogous to the setup in~\cite{Rey:1998bq, Brandhuber:1998bs}.
The equation \eqref{ENEN} represents the quark anti-quark potential. }

\begin{figure}[htbp]
     \begin{center}
   \includegraphics[height=4cm,clip]{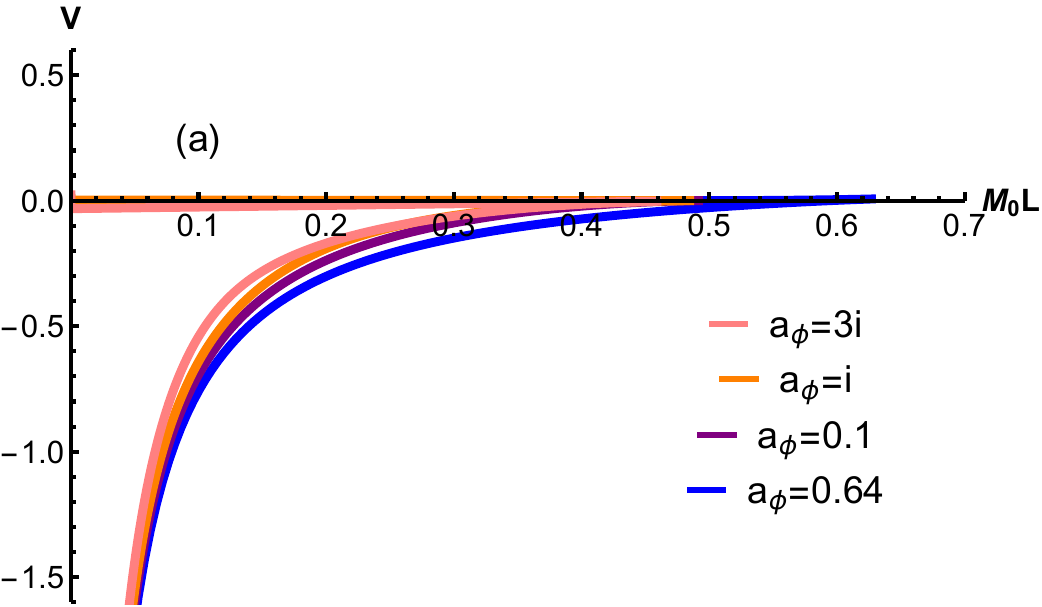}
   \includegraphics[height=4cm,clip]{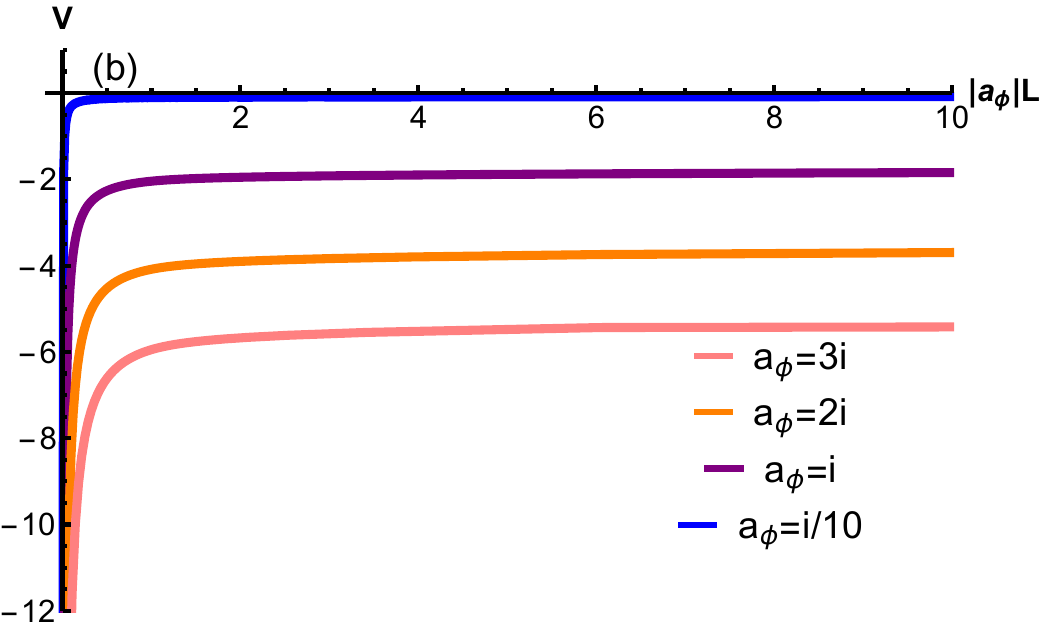}
           \caption{(a) The normalized quark anti-quark potential $V=\alpha'E/R^2$ is plotted as a function of $M_0L$ ($M_0=0.4$) with varying  $a_{\phi}$. Because the point of the zero potential shifts to large $L$  for large values of $a_{\phi}$, phenomena analogous to the dissociation are harder to occur.  (b) The quark anti-quark potential $V$ is shown as a function of $|a_{\phi}|L$ ($M_0=0$). Note that $a_{\phi}$ must be imaginary.}
    \label{fig:Wil2}
    \end{center}
\end{figure}

\begin{figure}[htbp]
     \begin{center}
   \includegraphics[height=5.5cm,clip]{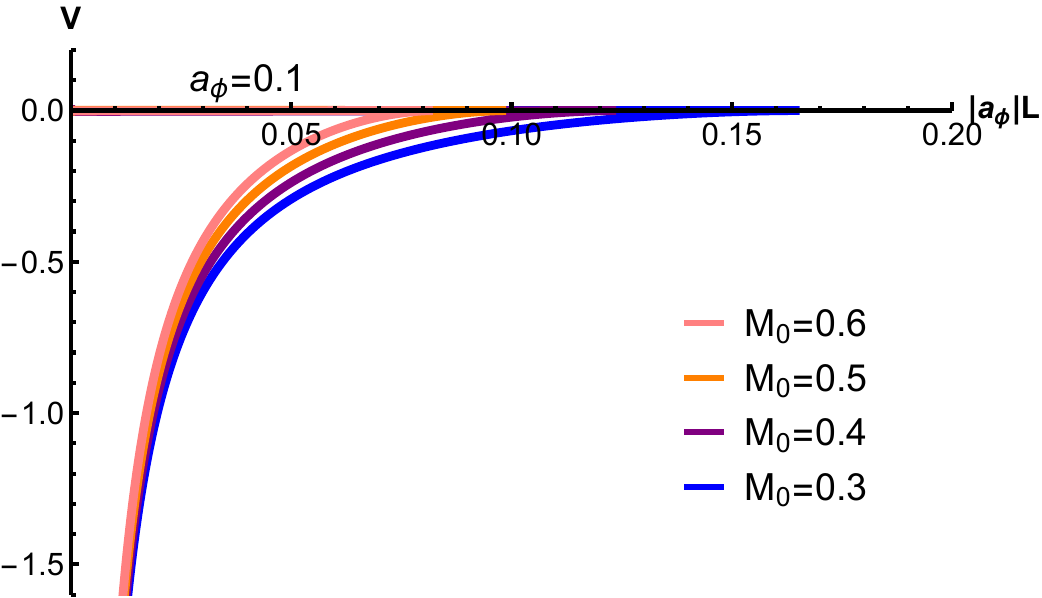}
           \caption{$M_0$ dependence of the normalized quark anti-quark potential is illustrated as a function of $M_0L$ ($a_{\phi}=0.1$). Because the zero-potential point shifts to large $L$ for small $M_0$, phenomena analogous to the dissociation are harder to occur.  }
    \label{fig:Wil3}
    \end{center}
\end{figure}
Fig. \ref{fig:Wil2} (a) illustrates $a_{\phi}$ dependence of the normalized quark anti-quark potential along $(\tau,\phi)$ directions. {The potential is Coulomb-like and possesses a finite range, analogous to~\cite{Rey:1998bq, Brandhuber:1998bs}.} The potential reaches zero at $L=L_c$, {which will result from the gravitational force.} Zero of the potential is analogous to the point of the dissociation of the heavy quarkonia. $L_c$ increases when $a_{\phi}$ increases, and physics analogous to the dissociation is less likely to occur. Moreover, the potential deepens with large values of $a_{\phi}$. It implies that the mass of heavy quarkonia decreases because the mass of heavy quarkonia is approximately equal to the sum of the masses of two heavy quarks plus the binding energy: $m_{Q\bar{Q}}\sim 2 m_Q+E_{Q\bar{Q}}$. When the Kaluza-Klein (KK) mass is zero, no phase transition occurs in Fig. \ref{fig:Wil2} (b). Due to scaling symmetry of EOM under $(t,\vec{x},z)\to \lambda^{-1} (t,\vec{x},z)$, parameters undergo a scaling transformation $(a_{\phi},V,1/L)\sim \lambda (a_{\phi},V,1/L)$. Consequently, the functional form of the potential should be $V=|a_{\phi}|F_2(|a_{\phi}|L)$.

Fig. \ref{fig:Wil3} depicts $M_0$ dependence of the normalized quark anti-quark potential along $(\tau,\phi)$ directions. The potential vanishes at $L=L_c$, indicating the point at which the dissociation-like phenomena occurs.  Because the point of the zero potential shifts to large $L$ for smaller values of $M_0$, physics analogous to the dissociation is less likely to occur.

\begin{figure}[htbp]
     \begin{center}
  \includegraphics[height=3.5cm,clip]{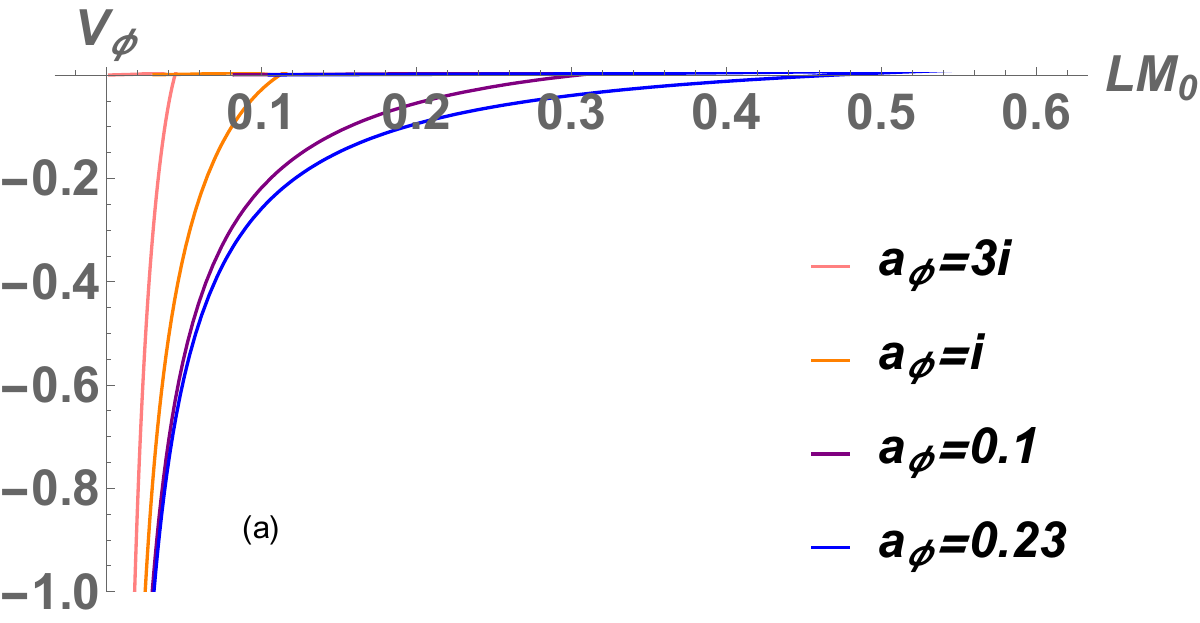}
   \includegraphics[height=3.5cm,clip]{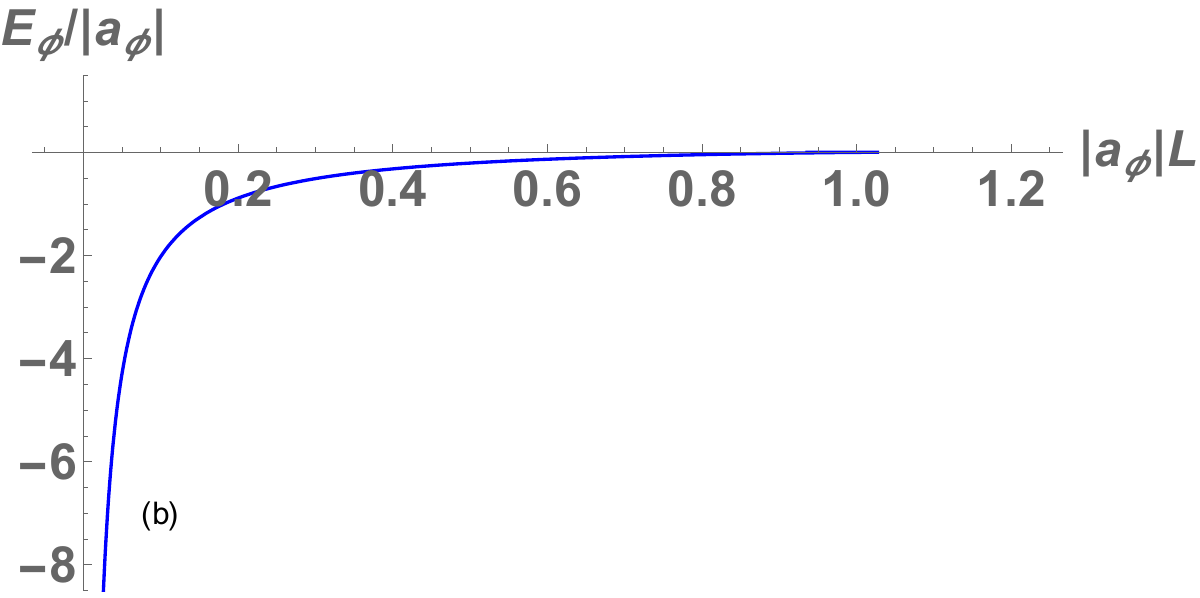}
           \caption{{(a) The normalized quark potential as a function of $LM_0$. The dissociation of quarkonium occurs when potential becomes zero. (b) The normalized quark potential as a function of $L|a_{\phi}|$. Unlike the case $z=z(\phi)$, dissociation occurs.} }
    \label{fig:Wilnew}
    \end{center}
\end{figure}

Note that the $\phi$ direction was originally the Euclidean time direction before the double Wick rotation. {This heavy quark potential is analogous to the one along $(\phi,\tau)$ directions, where $\phi$ represents Euclidean time and $\tau$ is a spatial direction ($-L/2\le \tau \le L/2$). Choosing the embedding scalar $z=z(\tau)$, the action is transformed to 
\ba
&S=\dfrac{1}{2\pi\alpha'}\int d\tau d\phi \sqrt{g_{\phi\phi}(g_{\tau\tau}+g_{zz}{(dz/d\tau )}^2)} \nonumber \\
&=\dfrac{R^2}{2\pi\alpha'}\int d\tau d\phi \sqrt{\dfrac{f(z)}{z^2}\Big(\dfrac{1}{z^2}+\dfrac{(dz/d\tau )^2}{z^2f(z)}\Big)}. 
\ea
Note that the string's worldsheet is oriented along the $(\tau,\phi)$ directions again. 
The inter-quark distance and regularized energy become
\ba\label{POT32}
&L=2\int_0^{z_0} dz \dfrac{ z^2 \sqrt{{f\left(z_0\right)}}}{(\sqrt{f(z)(z_0^4 f(z)- z^4 f\left(z_0\right)})}, \nonumber \\
&E_{\phi}=\dfrac{R^2}{\pi\alpha'}\int_0^{z_0} dz \left(\dfrac{z_0^2\sqrt{f(z)}}{z^2 \sqrt{z_0^4f(z)-{z^4 f\left(z_0\right)}}}-\dfrac{1}{z^2}\right)-\dfrac{R^2}{\pi\alpha'}\int^{z_+}_{z_0}\dfrac{dz}{z^2}.
\ea
Fig. \ref{fig:Wilnew}(a) shows the normalized quark potential as a function of $LM_0$. We observe dissociation of quarkonium at a point where potential becomes zero. As the temperature $M_0$ associated with it decreases, the potential deepens, resulting in a decrease in the mass of the heavy quarkonia. The dissociation-like phenomena occur quickly when the potential zero point $(L=L_c)$ decreases. This finding is consistent with the results in \cite{Rey:1998bq, Brandhuber:1998bs, Kim:2008ax}. Fig. \ref{fig:Wilnew}(b) shows the extremal limit of a quark potential. The dissociation occurs at $L|a_{\phi}|\sim 1$ even for the extremal limit ($M_0=0$) unlike \eqref{ENEN}. It implies that massive modes still cause phase transition in the extremal limit. The functional form of the potential is of the Coulomb type $\bar{V}\sim 0.77a_{\phi}-0.71/L$. In conclusion, Fig. \ref{fig:Wil3} aligns with the observation that massive modes decouple from others in the low energy limit.

\begin{table}
\caption[]{The mass and binding energy of Bottomonium are caluculated for $d=5$ and $M_0=0.145$ in units of GeV with respect to the potential \eqref{ENEN}. Parameters including $L_a^{-1}$ are expressed in units of GeV. The mass of the bottom quark is given as 4.803 GeV. }
\label{Tablemass}
\begin{center}
\begin{tabular}{cccc}
\toprule
$a_{\phi}$ & $L_a$ & Meson mass  &  Binding energy  \hfil \\
\midrule
$3i$ & 0.77  & 9.1  & -0.27  \\[5pt]
$i$ & 0.68 & 9.3  & -0.15 \\[5pt]
0.1  & 0.66 & 9.0 & -0.30 \\[5pt]
0.23 & 0.66  & 8.9 & -0.36 \\  \bottomrule
\end{tabular}
\end{center}
\end{table}

\begin{table}
\caption[]{Binding energy of Bottomonium is evaluated for $d=5$ and $M_0=0$ with respect to the potential \eqref{ENEN}. All parameters, including the inverse length scale $L_a^{-1}$, are written in units of GeV. The mass of the bottom quark is 4.803 GeV. }
\label{Tableext}
\begin{center}
\begin{tabular}{cccc}
\toprule
$a_{\phi}$ & $L_a$  & Meson mass  &  Binding energy  \hfil \\
\midrule
$0.1i$ & 2.1  & 9.3 & -0.13  \\[5pt]
$i$ & 2.0 & 5.8 & -1.9 \\[5pt]
$2i$ & 2.2 & 2.2 & -3.7 \\[5pt]
$3i$ & 1.8  & & -5.3 \\  \bottomrule
\end{tabular}
\end{center}
\end{table}

We consider a $5d$ heavy-quark system, which obeys the potential \eqref{ENEN} and \eqref{POT32}, and analyze heavy quarkonium. The potential \eqref{ENEN} and \eqref{POT32} are given by $5d$ heavy quarks probing extra-dimensions and $5d$ heavy quarks in $4d$ space inside $5d$, respectively. We also asssume that the potential has rotational symmetry in spatial directions.} We approximate the Schr\"{o}dinger equation to estimate the mass of heavy quarkonium.  The interactions between heavy quarks are characterized by quark anti-quark potential $\bar{V}(L)=\pi\alpha' E/R^2$ derived from the holographic model \eqref{ENEN} and \eqref{POT32}. The normalization is consistent with~\cite{Kim:2008ax}. These heavy quarks exhibit non-relativistic motion.

We employ an approximation method to obtain the mass of heavy quarkonium. Following the method outlined in~\cite{Ikhdair:2009fn}, one obtains the binding energy in the leading order. Quark anti-quark potentials in this method must satisfy the following criteria:
\ba
&\bar{V}(L)=-\gamma L^{-a}+\kappa L^b +v_0,\quad a,b>0.
&\bar{V}'(L)>0,\quad \bar{V}''(L)\le 0,
\ea
where $\gamma$ and $\kappa$ are positive constants and $v_0$ is a constant. The potential for $x=const$ \eqref{ENEN} meets these conditions. The binding energy in the leading order is found to be 
\ba\label{BIN32}
E_{n,l}\sim \bar{V}(L_a)+\dfrac{1}{2}L_a \bar{V}'(L_a).
\ea
The parameter $L_a$ is determined by
{\ba
d-3+2l+(2n_R+1)\Big(3+\dfrac{r_a \bar{V}''(L_a)}{\bar{V}'(L_a)}\Big)^\frac{1}{2}=(8\mu L_a^3 \bar{V}'(L_a))^{\frac{1}{2}}
\ea}
where $n_R,\ l=0,1,2,\dots$ and $\mu$ denote the reduced mass.

The mass of the bound state is given by
\ba\label{PAR33}
m_{\bar{Q}Q}=2m_Q+2 E_{n,l},
\ea
{where $m_Q$ is the mass of a heavy quark.}

{We assume that the input of a quark mass is a bottom quark mass $m_Q=4.803$ GeV, while this is the mass of a $5d$ heavy quark.} For a bottom-quark system, the reduced mass is given by  $\mu=m_1m_2/(m_1+m_2)=m_Q/2$.  The approximation of a non-relativistic motion is good because the mass of a bottom quark 4.8 GeV is much larger than $M_0=0.145$ GeV. By substituting this into \eqref{BIN32} and \eqref{PAR33}, one obtains the binding energy and $L_a$ for the potential \eqref{ENEN}. See Table. \ref{Tablemass}. 
Because the binding energy is negative, a bound state, namely, bottomonium is formed.~\footnote{For a charm-quark system, the bound energy becomes positive. It indicates that no bound state is formed for $M_0=0.4$ GeV. This is due to the value of $M_0$. For small $M_0$, charmonium will be formed.} Bottomonium mass decreases with increasing $a_{\phi}^2$ (energy) as expected. {Note that Bottomonium states $\Upsilon(1S)$ and $\eta_b(1S)$ have the masses 9.46 GeV and  9.40 GeV. The mass of holographic heavy quarkonium is slightly smaller than that of bottomonium. }

%For a bottom-quark system, a bound state, namely, bottomoinum is formed.~\footnote{For a charm-quark system, the bound energy becomes positive. It shows that the bound state is not formed.} Mass of bottomoinum decreases with increase of $a_{\phi}^2$ (energy) as expected. See Table. \ref{Tablemass}.

In the extremal limit, the mass of bottomonium decreases because the binding energy diminishes and even approaches to zero. However, we cannot definitively conclude the zero mass of bottomonium because of the approximations involved.  The functional form of the potential is of the Coulomb type $\bar{V}\sim -0.22/L+c_v|a_{\phi}|$.

{
\begin{table}
\caption[]{The mass and binding energy of Bottomonium are caluculated for $d=5$ and $M_0=0.145$ GeV with respect to \eqref{POT32}. Parameters including $L_a^{-1}$ are expressed in units of GeV. The mass of the bottom quark is given as 4.803 GeV. }
\label{Tablemasstau}
\begin{center}
\begin{tabular}{cccc}
\toprule
$a_{\phi}$ & $L_a$ & Meson mass  &  Binding energy  \hfil \\
\midrule
$3i$ & 0.91  & 12.52  & 1.46  \\[5pt]
$i$ & 0.67 & 10.42  &  0.41 \\[5pt]
0.1  & 0.66 & 9.2 & -0.20 \\[5pt]
0.23 & 0.66  & 9.0 & -0.32 \\  \bottomrule
\end{tabular}
\end{center}
\end{table}
We then consider quark potential in \eqref{POT32}. Binding energy and bottomonium mass are given in Table. \ref{Tablemasstau}. Binding energy of bottom quarks changes from a positive value to a negative value. Thus, bottomonium is formed. For high temperature $M_0$, on the other hand, binding energy becomes positive and the bound state is not formed. Results in the extremal limit are similar. 
}

\section{Massless dilaton}\label{sec3}
In this section, we calculate a holographic mass spectrum of the glueball-like operator dual to the massless dilaton in general dimensions.~\footnote{In QCD, the scalar glueball corresponds to the operator $\text{tr} F_{\mu\nu}F^{\mu\nu}$.} {While in $4d$ $\mathcal{N}=4$ SYM, this operator is a supersymmetric completion of  $\text{tr} F_{\mu\nu}F^{\mu\nu}$. However, supersymmetry is broken due to the boundary condition and quantum corrections in the QFT dual to \eqref{MET01}. We focus on modes independent of $\tau$ and the following Fourier component $\Phi =\phi (r)e^{i kx}$~\cite{Csaki:1998qr}.} {Using \eqref{MET01}, the EOM of the massless dilaton become
\ba\label{proschodingereq}
\partial_z \Big(\dfrac{f(z)}{z^{d-1}}\partial_z\phi (z)\Big)-\dfrac{k^2}{z^{d-1}}\phi (z)=0.
\ea}\footnote{When we perform the coordinate transformation $r=1/z$, we arrive at the following expression: \ba
\dfrac{1}{\sqrt{-g}}\partial_\mu \sqrt{-g}\partial^\mu \phi )=\dfrac{1}{r^3}\partial_r (r^5 f_2(r)\partial_r \phi)-\dfrac{k^2}{r^2}\phi =0,
\ea
where $f_2(r)=1-\Big(1-\frac{a_\phi ^2}{r_h^2}\Big)\Big(\frac{r_h}{r}\Big)^d-\Big(\frac{a_\phi}{r_h}\Big)^2 \Big(\frac{r_h}{r}\Big)^{2d-2}$. }

\begin{table}
\caption[]{$0^{++}$ glueball spectrum in 3 dimensional QCD with gauge potential. The mass is expressed in units of the square root of the string tension.}
\label{Tableglue4}
\begin{center}
\begin{tabular}{ccccc}
\toprule
States & Lattice QCD   & $a_{\phi}=0,\ M_0=0.405$ & $a_{\phi}=2\pi M_0/3,\ M_0=0.571$  \hfil \\
\midrule
$0^{++}$ & 4.33  & 4.33 & 4.33 \\[5pt]
$0^{++*}$ & 6.52  & 7.48 & 7.56 \\[5pt]
$0^{++**}$ & 8.23  & 10.6 & 10.7 \\[5pt]
$0^{++***}$ &  & 13.6 & 13.8 \\  \bottomrule
\end{tabular}
\end{center}
\end{table}
\begin{table}
\caption[]{$0^{++}$ glueball spectrum in 4 dimensional QCD with gauge potential. The mass is in units of GeV. Parentheses mean anisotropic ones.}
\label{Tableglue5}
\begin{center}
\begin{tabular}{ccccc}
\toprule
States & Lattice QCD   & $a_{\phi}=0$   & $a_{\phi}=\pi M_0/2$ &$a_{\phi}=(7 i)/75$ \hfil \\
 &   &  $M_0=0.145$ &  $M_0=0.218$ &  $M_0=0.140$ \hfil \\
\midrule
$0^{++}$ & 1.48-1.58(1.73) &  1.48 & 1.48 &1.48  \\[5pt]
$0^{++*}$ & 2.76-2.84(2.67) & 2.43 & 2.46 &2.44 \\[5pt]
$0^{++**}$ & 3.37 &  3.36 &3.41&3.37\\[5pt]
$0^{++***}$ & 3.99  & 4.27 & 4.35&4.29 \\  \bottomrule
\end{tabular}
\end{center}
\end{table}

\begin{table}\caption[]{$0^{++}$ glueball spectrum in 4 dimensional QCD with gauge potential. The mass is in units of GeV. Parentheses mean anisotropic cases.}
\label{Tableglue6}
\begin{center}
\begin{tabular}{ccccc}
\toprule
States & Lattice QCD   & $a_{\phi}=0,\ M_0=0.170$ & $a_{\phi}=i/5,\ 
 M_0=0.150$  \hfil \\
\midrule
$0^{++}$ & 1.48-1.58(1.73) &  1.73 & 1.73   \\[5pt]
$0^{++*}$ & 2.76-2.84(2.67) & 2.85 & 2.84  \\[5pt]
$0^{++**}$ & 3.37 &  3.94 &3.93  \\[5pt]
$0^{++***}$ & 3.99  & 5.02 & 5.00 \\  \bottomrule
\end{tabular}
\end{center}
\end{table}

{The asymptotic behaviors of the field $\phi$ are $\phi\sim 1,\ z^{d}$ near the AdS boundary, which implies that the dimension of the corresponding operator is $\Delta =d$. } The asymptotic behavior at the tip includes the log divergences.~\footnote{For $d=4$, the asymptotic expansion at the tip is $\phi \sim a_1+a_2 \log (z-1/r_h)-a_2 k^2 r_h/(2 (a_{\phi}^2+2 r_h^2))(z-1/r_h)\log (z-1/r_h)\dots$, where $r_h=1/z_+$.} Due to this divergence at $z=z_+$, we impose a regular boundary condition at the tip. We employ a shooting method to determine the glueball mass. The glueball mass corresponds to an eigenvalue of $-k^2$. By setting $k^2=-m^2$, one can obtain numerical solutions and match them with the asymptotic values at the AdS boundary. We require that $\phi$ vanishes at the AdS boundary. 

One can also employ a Schr\"{o}dinger-like equation for an eigenvalue problem. {By applying a Bogoliubov transformation $\psi(z)=\phi(z)/\sqrt{h(z)}$, where $h(z)=f(z)/z^{(d-1)}$, Eq. (\ref{proschodingereq}) can be transformed into a Schr\"{o}dinger like equation
\ba
-\psi''(z)+V(z)\psi(z)=-k^2\psi(z),
\ea
with the potential
\ba
V(z)=-\frac{-\left(d^2-1\right) f(z)^2-2 z f(z) \left(z \left(f''(z)+2 k^2\right)-(d-1) f'(z)\right)+z^2 f'(z)^2+4 k^2 z^2 f(z)^2}{4 z^2 f(z)^2}
\ea}
Because potential $V(z)$ depends on both the momentum $k^2$ and $f(z)$, its form of the potential varies with $k^2$. Fixing $M_0$ and $k^2$, the potential becomes deeper and $z_+$ increases when $a_{\phi}$ increases.

For $d=4$, we set values $M_0=\frac{1}{\pi}$. The spectrum of the glueball-like operator then becomes
\ba
&m^2=11.6,\quad 34.5,\quad 69.0,\quad 115,\dots \quad & \text{for $a_\phi =0$}, \\
&m^2=3.60,\quad 11.1,\quad 22.4,\quad 37.3,\dots \quad & \text{for $a_\phi =\frac{1}{\sqrt{2}}$}, \\
&m^2=19.8,\quad 58,\quad  115,\quad 192,\dots \quad & \text{for $a_\phi =i$}.
\ea
It indicates that the mass decreases with the increase of squared gauge potential $a_{\phi}^2$ as anticipated from the computation of the generalized entropic C-function~\cite{Fujita:2020qvp, Fujita:2023bdk}. Recall that pure imaginary $a_{\phi}$ corresponds not to an imaginary chemical potential but instead to a chemical potential in the original Reissner Nordstr\"{o}m AdS black hole. The extremal case $M_0=0$ differs from the case with non-zero $M_0$. Because $f'(z_+)\propto (d+(d-2)z_+^2a_{\phi}^2)=0$, the asymptotic expansion at the tip becomes divergent and changes for the extremal case. Consequently, glueball-like operators are massless. 

Results of the gravity dual are compared with the spectrum of spin $J^{PC}=0^{++}$ glueball-like operators for QCD$_3$ in Table \ref{Tableglue4}, where the spin is denoted by $J$, while $P, C$ represent the parity and charge conjugation quantum numbers, respectively.  Results of lattice QCD are sourced from \cite{Teper:1998te}. The mass values are expressed in units of the square root of the string tension. The Kaluza-Klein mass is selected based on the input mass of $0^{++}$ states is input. The fourth column is for critical gauge potential (with zero boundary energy). Because the free energy of the AdS black hole is negative, the AdS soliton with gauge potential will be unstable at a finite temperature in this scenario. The spectrum closely resembles the case of zero gauge potential while the Kaluza-Klein mass $M_0$ increases.

%When $M_0=0$ (the extremal case), the spectrum of the glueball-like operator becomes 
%\ba
%&m^2=0.52,\quad 0.94,\quad 1.60,\quad 2.49,\dots\quad &  \text{for $a_\phi =i/2$}, \\
%&m^2=1.94, \quad 3.21,\quad 5.32,\quad 8.27,\dots \quad & \text{for $a_\phi =i$}, \\
%&m^2=4.75,\quad 9.80,\quad 18.2,\quad 30.9,\dots \quad & \text{for $a_\phi =2i$}, 
%\ea

For $d=5$, the spectrum of the glueball-like operator is also obtained. The KK mass $M_0$ is fixed to be 0.4 in units of GeV. The resulting  spectrum of the glueball-like operator becomes
\ba
&m^2=30.8,\quad 82.3,\quad 157,\quad 254,\dots &  \text{for $a_\phi =i$}, \\
&m^2=16.7,\quad 45.2,\quad 86.5,\quad 140,\dots &  \text{for $a_\phi =0$}, \\
&m^2=7.35,\quad 20.4,\quad 39.1,\quad 63.5,\dots &  \text{for $a_\phi =\pi M_0/2$}, \\
&m^2=5.07,\quad 14.2,\quad 27.2,\quad 44.4,\dots & \text{for $a_\phi =2\pi M_0/\sqrt{15}$}.
\ea
As expected, the mass decreases with the increase of $a_{\phi}^2$. Therefore, the glueball-like operator contributes to degrees of freedom at lower energy. The extremal case $M_0=0$ differs significantly. Glueball-like operators become massless in that case. Recall that the original extremal black hole has a chemical potential instead of an imaginary chemical potential, and one can not activate positive $a_{\phi}$ for the extremal case.

The results from the gravity dual are compared with the spectrum of spin $0^{++}$ glueball-like operators for QCD$_4$, as detailed in Table \ref{Tableglue5} and \ref{Tableglue6}. The second column lists the glueball masses from lattice QCD, in units of GeV, as reported in \cite{Li:2013oda, Zhang:2021itx}. The range reflects several datasets. Parentheses mean anisotropic ones~\cite{Meyer:2004gx, Morningstar:1999rf}. The Kaluza-Klein mass is adjusted to reproduce the $0^{++}$ states. The critical gauge potential is specified in the fourth column, and the energy $M$ is set to zero. The spectrum resembles the case of a zero gauge potential, except for an increase in the Kaluza-Klein mass. The Kaluza-Klein mass $M_0 = 0.218$ GeV is nearly equal to the QCD critical temperature $T_c = 0.28$ GeV and the $\Lambda$-scale $\Lambda_{MS} = 0.25$ GeV~\cite{Itou:2015cyu}. However, due to thermodynamic instability, the AdS soliton with a gauge potential becomes unstable at finite temperatures.

%When $M_0=0$ (the extremal case), the spectrum of the glueball-like operator becomes 
%\ba
%&m^2=1,\quad 1.69,\quad 2.79,\quad 4.2,\dots\quad &  \text{for $a_\phi =i/2$}, \\
%&m^2=3.38, \quad 5.27,\quad 8.97,\quad 14.4,\dots \quad & \text{for $a_\phi =i$}, \\
%&m^2=5.50,\quad 18.5,\quad 34.4,\quad 57.8 \dots  \quad & \text{for $a_\phi =2i$}, 
%\ea

\section{Discussion}

We investigated holographic heavy quark potential from holographic Wilson loops in the AdS soliton with gauge potential. We examined two kinds of loops. For $\phi =$const, holographic heavy quark potential exhibits the area law behavior {as a function of the inter-quark distance.} The QCD string tension (the slope of the potential) decreases with an increase in gauge potential. It implies that the mass of the excitations decreases, as expected from the results of holographic renormalized entanglement entropy and the entropic C-function~\cite{Fujita:2023bdk}. In the extremal limit ($M_0=0$), we still observe the area law for $a_{\phi}\neq 0$, while Kaluza-Klein modes become massless. It shows that the mass of excitations is nonzero. The QCD string tension increases by $|a_{\phi}|$. 

{Note that Hawking-Page phase transition will affect heavy quark potential. The AdS black hole is more stable at large gauge potential $a_{\phi}>a_{\phi, c}$ and Hawking-Page phase transition occurs because free energy of the AdS soliton with a gauge potential even becomes positive at that case. The area law will be valid for small $a_{\phi}$, while it disappears under conditions $a_{\phi}\ge a_{\phi, c}$, where quark anti-quark potential disappears, showing deconfinement. Therefore, the twisting parameter prefers anti-periodic boundary conditions on fermions, leading to chiral symmetry breaking and then confinement at $M_0L >>1$. }

For $x=$const and non-zero Kaluza-Klein mass, physics analogous to the dissociation occurs. {the potential has a finite range and vanishes at a critical inter-quark distance $L=L_c$}. 
 When $a_{\phi}$ increases, physics analogous to the dissociation is less likely to occur. The potential deepens with an increase of $a_{\phi}$. It implies that the mass of heavy quarkonia decreases due to the relation $m_{Q\bar{Q}}\sim 2m_Q+E_{n,l}$, where $m_Q$ is the mass of a heavy quark and $E_{n,l}$ is binding energy as shown in Table \ref{Tablemass}. The mass of holographic heavy quarkonium is less than the mass of bottomonium a little. In the extremal limit ($M_0=0$), the Kaluza-Klein mass is zero, and no phase transition occurs. Table \ref{Tableext} shows that even the meson mass approaches zero for large $|a_{\phi}|$.

%When Kaluza-Klein mass is non-zero, physics analogous to the dissociation occurs.

We have also calculated the mass of glueball-like operators dual to massless dilaton in $4$ and $5$ dimensions. The mass of $0^{++}$ glueball-like operator decreases with the increase of $a_{\phi}$ as anticipated. Our results correspond to modified boundary conditions and are comparable with those from lattice QCD. The mass of the third and fourth excited states in the anisotropic cases has also been obtained holographically. Interestingly, we discovered that choosing appropriate values for the gauge potential in our holographic model can make the predicted spectrum even closer to that computed using lattice QCD. This investigation may provide deeper insights into the AdS/QCD correspondence.} 

Adding the antisymmetric tensor $B$ is also another  direction~\cite{Ghoroku:2020fkv}. It is essential for ensuring the periodicity of $a_{\phi}$. For finite $N_c$, large values of $a_{\phi}$ are allowed. Additionally, the system is stabilized in the presence of $B$. It would be valuable to perform a double wick rotation of a QFT with an imaginary chemical potential and demonstrate this periodicity using both the QFT side and the path integral method.

\section*{Acknowledgments}
MF would like to thank Takeshi Morita for their helpful discussion. We would like to thank Song He for collaboration in the initial stage of this project and for helpful discussions. This work is supported in part by the National Natural Science Foundation of China (NSFC) Grant Nos: 12405154 and the European Union -- Next Generation EU through the research grant number P2022Z4P4B ``SOPHYA - Sustainable Optimised PHYsics Algorithms: fundamental physics to build an advanced society'' under the program PRIN 2022 PNRR of the Italian Ministero dell'Universit\`a e Ricerca (MUR).. M.F. and B.C. thank SCNT, Institute of Modern Physics, for their hospitality. 

\appendix

\section{Background Wilson lines}\label{appA}
This section examines the open string spectrum in the presence of flat background gauge fields $a_{\phi}$. This background field gives rise to non-trivial Wilson lines. The constant gauge field is considered. It can be diagonalized by using gauge transformations:
\ba
a_{\phi}=\mbox{diag}(a_{\phi}^{(1)},\a_{\phi}^{(2)},\dots , a_{\phi}^{(N)}).
\ea
This configuration represents a pure gauge and has vanishing field strength. 
Here, the non-Abelian gauge transformation is defined as
\ba
A_{\mu}\to UA_{\mu}U^{-1}+i U\partial_{\mu}U^{-1},
\ea
and then
\ba
D_{\mu}\to UD_{\mu}U^{-1},\quad F_{\mu\nu}=i[D_{\mu},D_{\nu}]\to UF_{\mu\nu}U^{-1},
\ea
The gauge transformation that sets $a_{\phi}$ to zero is given by
\ba\label{UNI23}
U=\mbox{diag}(e^{-ia_{\phi}^{(1)}\phi },e^{-ia_{\phi}^{(2)}\phi},\dots , e^{-ia_{\phi}^{(N)}\phi}).
\ea
The non-Abelian field in the adjoint representation $X$ is transformed under \eqref{UNI23} as
\ba
X\to UXU^{-1},\quad X_{ij}\to e^{-i (a_{\phi}^{(i)}-a_{\phi}^{(j)})\phi}X_{ij}.
\ea
While the original field $X$ on the left-hand side is periodic along 
 the $\phi$ direction with KK momentum $p_{\phi}^{(0)}=2\pi nM_{0}$, the transformed field $X$ does not need to remain periodic along the compactified direction.  
Instead, the transformed field $X$ acquires a shifted momentum along the $\phi$ direction as follows:
\ba
p_{\phi}=2\pi n M_0-a_{\phi}^{(i)}+a_{\phi}^{(j)},
\ea
where the first term corresponds to KK momentum along $\phi$ characterized by an integer $n$.

Consequently, the open string spectrum becomes
\ba\label{OPE25}
m^2=\Big(2\pi n M_0-a_{\phi}^{(i)}+a_{\phi}^{(j)}\Big)^2+\dfrac{1}{\alpha'}(N_o-1).
\ea
In a general background, all $a_{\phi}^{i}$ values are distinct, and $i =j$ modes don't depend on Wilson lines. For massless gauge field, $N_o=1$ and $n=0$. $i=j$ modes are massless modes. In that case, the unbroken gauge symmetry is $U(1)^N$. If $r$ of the $a_{\phi}^{(i)}$ values are equal, $r\times r$ gauge fields become massless, carrying $U(r)$ gauge symmetry.

This open string spectrum \eqref{OPE25} indicates that if the difference $\Delta a_{\phi}=a_{\phi}^{(i)}-a_{\phi}^{(j)}$ increases, the squared mass becomes zero when $\Delta a_{\phi}=2\pi nM_0$ and $N_o=1$. This mode originates from the compactified direction. Because the gauge potential decreases the mass, the open string spectrum resembles the glueball spectrum discussed in section \ref{sec3}.

\end{document}